\documentclass{cernrep}

\usepackage{amsfonts}
\usepackage{amssymb}
\usepackage{amsmath}
\usepackage{mathrsfs}
\usepackage{fancyhdr}
\fancyhfoffset{4 mm}
\fancypagestyle{ARTTITLE}{%
\fancyhf{} 
\lhead{\small{Proceedings of the 2018 CERN--Accelerator--School course on\\
\it{Numerical Methods for Analysis, Design and Modelling of Particle Accelerators}, Thessaloniki, (Greece)}} 
\lfoot{Available online at \url{https://cas.web.cern.ch/previous-schools}}
\rfoot{\thepage\hspace*{3mm}}
 
}

\begin{document}

\addtolength{\topmargin}{0.5cm}
\addtolength{\evensidemargin}{-0.5cm}
\setlength{\voffset}{0cm}

 \title{Measuring the Field Quality in Accelerator Magnets with \newline the Oscillating-Wire Method - a Case Study for Solving Partial Differential Equations} \author{Stephan Russenschuck}
\institute{CERN, 1211, Geneva 23, Switzerland}
\maketitle
\thispagestyle{ARTTITLE}

\newcommand{\bq  }{\begin{equation}}
\newcommand{\eq  }{\end{equation}}
\newcommand{\bqn  }{\begin{align}}
\newcommand{\eqn  }{\end{align}}
\newcommand{\russ}{\hspace{.5in}}
\newcommand{\riss}{\hspace{.2in}}

\newcommand{\curv}{{\mathscr S}}
\newcommand{\surf}{{\mathscr A}}
\newcommand{\vol}{{\mathscr V}}
\newcommand{\curvc}{{\mathscr C}}
\newcommand{\spacel}{{\mathscr M}}
\newcommand{\spacll}{{\mathscr N}}
\newcommand{\domain}{{\mathscr D}}
\newcommand{\n } {\mathrm}
\newcommand{\ttd } {\mathrm{d}}
\newcommand{\td } {\mathrm{d}}

\newcommand{\curvl}{{\mathscr L}}
\newcommand{\point}{{\mathscr P}}
\newcommand{\origin}{{\mathscr O}}
\newcommand{\pointl}{{\mathscr Q}}
\newcommand{\pointb}{{\mathscr B}}
\newcommand{\disc}{{\mathscr D}}

\renewcommand{\div}{ \, \mathrm{div} \,}
\newcommand{\dif}{ \, \mathrm{div} \,}
\newcommand{\curl}{ \, \mathrm{curl} \,}
\newcommand{\grad}{ \, \mathrm{grad} \,}
\newcommand{\expec}[1]{\mathbb{E}\left[#1\right]}
\renewcommand{\d}{\text{d}}

\begin{abstract}
The single stretched-wire method is commonly used to measure the magnetic field strength and magnetic axis in an accelerator magnet. The integrated voltage at the connection terminals of the wire is a measure for the flux linked with the surface traced out by the displaced wire. The stretched wire can also be excited with an alternating current well below the resonance frequency. It is thus possible to measure multipole field errors by making use of the linear relationship between the wire-oscillation amplitude, integrated field, and current amplitude. \par

This technique is a good example for solving partial differential equations, or more precisely, boundary value problems in one and two dimensions. In particular, the field in the aperture of accelerator magnets is governed by the~Laplace equation, which leads to a boundary-value problem that is solved by determining the coefficients in the series of eigenfunctions from measurements of the~field components or wire-oscillation amplitudes on the domain boundary. 

The oscillation of the taut string is an example of a one-dimensional, in-homo- genous wave equation. The metrological characterization of the oscillating-wire system yield ~the feedback on the uncertainties (and limitations) ~of the~method, as only the linearized equations of the wire motion and the integrated field harmonics of the magnet are considered.
\end{abstract}

\section{Introduction}

It is said that (partial) differential equations are {\em solved} employing specific, albeit not universal, methods, of which there are far more than can be discussed in this short writeup. 
However, {\em we}, engineers and applied physicists, are usually not solving partial differential equations; as they have been solved for us, we look them up in a book. \par

What we {\em solve} indeed are boundary-value problems in a coordinate system appropriate to the~problem domain, for example, the round or elliptical aperture of an accelerator magnet. 
For this purpose we look up functions that obey the differential equations that are often been derived after serious linearization and only for trivial domains. In any other case, one has to resort to numerical methods. \par

If said functions are orthogonal and complete, a finite sum over these eigenfunctions can be shown to be the best approximation of a systems response in this dimension. The sum of eigenfunctions must then be matched with given initial and/or boundary data derived from mere assumptions, computations, or measurements. 

In the aperture of an accelerator magnet, free of currents and magnetized material, both magnetic scalar and vector potentials can be employed for the formulation of a boundary value problem. For magnets that 
are long compared to their fringe-field region, a 2D field description of the integrated field quantities is often sufficient. The domain boundary is often chosen as a circle
with a outer radius of two-thirds of the aperture radius. Both potential formulations yield scalar Laplace equations. Its eigenfunctions are the trigonometric functions and thus the field quality is conveniently described by a~set of coefficients, known as field harmonics.

These coefficients are then determined by comparison with the boundary values; in circular coordinates these boundary values are given by the radial or azimuthal field components (or the horizontal or vertical wire-oscillation amplitudes) that are developed into Fourier series. If magnet design is not in our agenda (as for beam-physicists or magnetic  measurement engineers), the problem will be solved.  

However, by studying the theorems on harmonic functions we will find that if a scalar field is harmonic
in a closed domain, then the potential cannot take a maximum or minimum at any interior point of that domain. Harmonic fields are therefore unable to
account for the line-currents used for the~calculation of coil fields in (superconducting) magnets. Through the superposition principle for linear operator problems, 
the convolution of a Green's function with an excitation function will also obey the~differential equation. If the linear differential operator admits a set of 
eigenvectors, i.e., eigenfunctions and scalar coefficients, then it is possible to develop the Green's function into these eigenfunctions, a process know as Fredholm theory. 

In this write-up, we will focus on the oscillating-wire technique for measuring magnetic fields in accelerator magnets. It is a fine case study for the treatment of boundary-value problems. The method requires the study of field harmonics in the aperture of the magnet and of the wave equation for a taut string acting as a transducer for the integrated magnetic field. 

\section{The architecture of the stretched-wire systems} 
\begin{figure}[b]
\centering
\includegraphics[clip=,width=11cm]{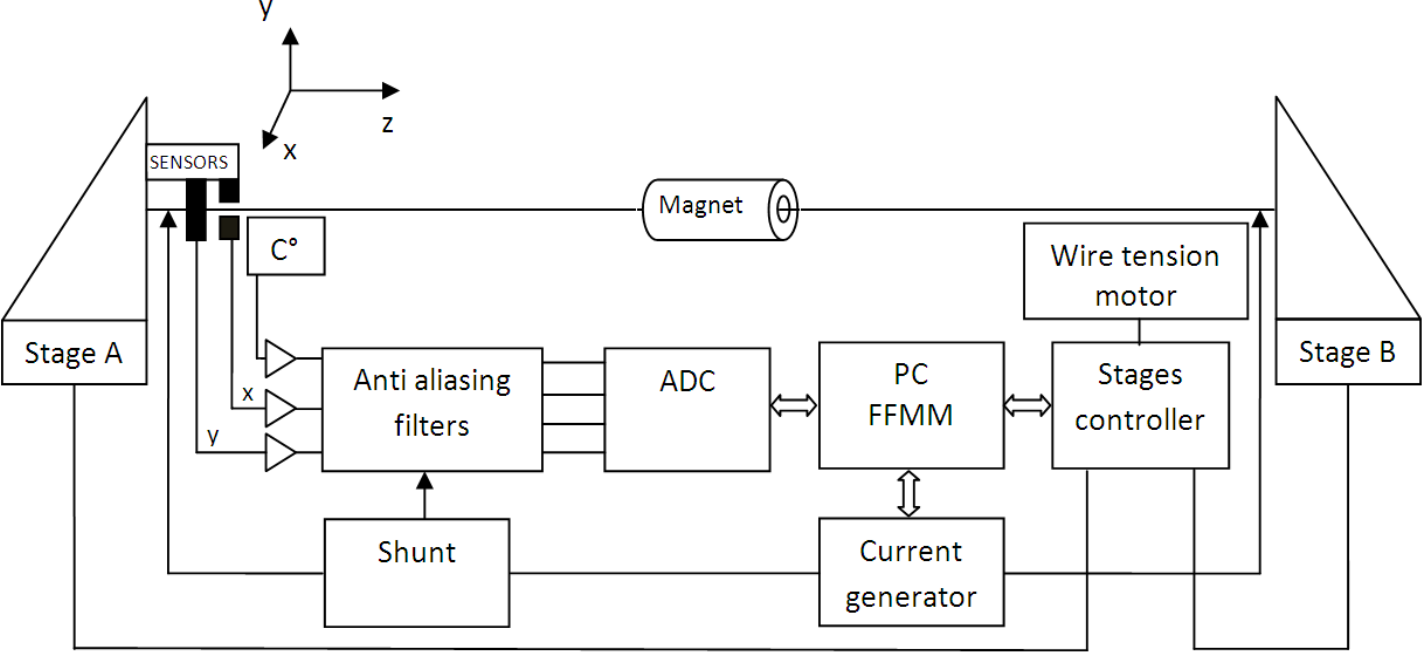}
   \caption{Systems architecture of the wire system.}
   \label{fig:SetUpArchitecture}
\end{figure}

Figure~\ref{fig:SetUpArchitecture} shows the systems architecture of the wire measurement system. A Copper-Beryllium wire of diameter $0.125$ mm is guided by two displacement stages (denoted A and B), at which the two end-points of the wire are fixed by ceramic ball bearings and kept electrically on a floating potential. 
The~wire can be considered free of martensitic contaminations, which would otherwise introduce a position-dependent magnetic force and add an uncertainty to the measurement. 
The wire is passed through the magnet aperture and pulled taut by means of a servo motor. \par
Two phototransistors are positioned on Stage A in order to transduce the wire oscillations in $x-$ and $y-$directions into voltages.  
This sensor output, pre-amplified and low-pass filtered by an anti-aliasing filter, is sent to a $18$ bit acquisition system. 
The same system also acquires the wire-excitation current on a reference resistor.  \par
The voltage integration is performed by the Fast Digital Integrator (FDI) V.3, a PXI-based card with an on-board digital signal processor. The integrator includes an analog front-end, based on a fully programmable input amplifier and features a self-calibration capability. The Flexible Framework for Magnetic Measurements (FFMM) is used as a software framework for data acquisition, driving the motor controllers and measurement
post-processing.  \par
\section{\label{taut}The vibration of a taut string}
The stretched wire can be described as a line structure, parametrized with respect to its arc length $s$ and 
suspended at its two extremities. This is often referred to as the {\em taut string}. If the displacement from the equilibrium position is small, 
the vibration is governed by linear differential equations. The classical analysis of these equations leads to the so-called normal modes of vibration. 
The following assumptions must be made, and shall be challenged at a later stage.
\begin{itemize}
\item Uniformity: The string has a constant linear mass density $\lambda_\n{m}$.
\item Rigid ends: There are no parametric excitations from end motions in transverse or axial directions.  
\item No bending stiffness: The string has no bending stiffness, that is,
the flexural and torsional rigidities can be neglected. 
\item Planar oscillations: The string deflection $u(z,t)$ is caused by the distributed force $f(z,t)$, which  is proportional to the local velocity and the normal component of the magnetic flux density to this $uz$-plane. 
\item Uniform tension: Each segment of the string pulls on its neighbouring segments with the same magnitude of force $T$.
\item The only force in the string is its tension and the Lorentz force exerted by the driving current and magnetic field. 
\item Small vibrations: The slope can be sufficiently well approximated by $\partial u(z,t) / \partial z$ on the entire interval [0, L].
\item Steady state oscillations: After an initial setting time, the string oscillates in the form of a standing wave. Then there will be no energy flow
along the string and no energy loss in the fixed suspensions.
\end{itemize}
Consider fixed points are at $z=0$ and $z=L$. Let the cross-sectional area of the wire be $A$. 
If there is an initial stretching of $\Delta L$, the initial tension is 
\bq T = E A \frac{\Delta L}{L} \, , \eq
according to Hook's law, where E is the Young modulus, [E] = 1 N m$^{-2}$, and T is the tension $[T] = 1 \, N = 1 \, \n{kg\, m \,s}^{-2}$.
Contrary to a beam, a string has no bending stiffness, that is,
the flexural and torsional rigidities can be neglected. When the deflections are small $(s \approx z)$, 
the distributed force is independent of the deflection and the tension remains the same along the string.\par
The tension varies in time, however. This is mainly due to the elongation of the wire at maximum displacement, 
uncertainties in the stage alignment, torque ripples in the tensioning motor,
and variations of the friction in the end-points. And even though the vibration amplitude is small, it is still large with respect 
to the sag. \par
The boundary values for the mathematical model are given by 
\bq u(0,t) = u(L,t) = 0 \eq 
and the initial displacement and transversal velocity 
are \bq u(z,0) = u_0(z), \russ \frac{\partial u }{ \partial t } (z,0) = v_0(z) \, . \eq
This is the characteristic of a Cauchy initial value problem. In the first class of initial conditions 
the~string is released from a displaced position with zero velocity. This is the condition of a plucked string on a guitar. In the second class of initial conditions
the string starts from the equilibrium position with specified velocities like in struck-string instruments such as the piano. \par
\begin{figure}[h]
\centering
\setlength{\unitlength}{1.0mm}
\begin{picture}(80,55)
\includegraphics[clip=,width=8cm]{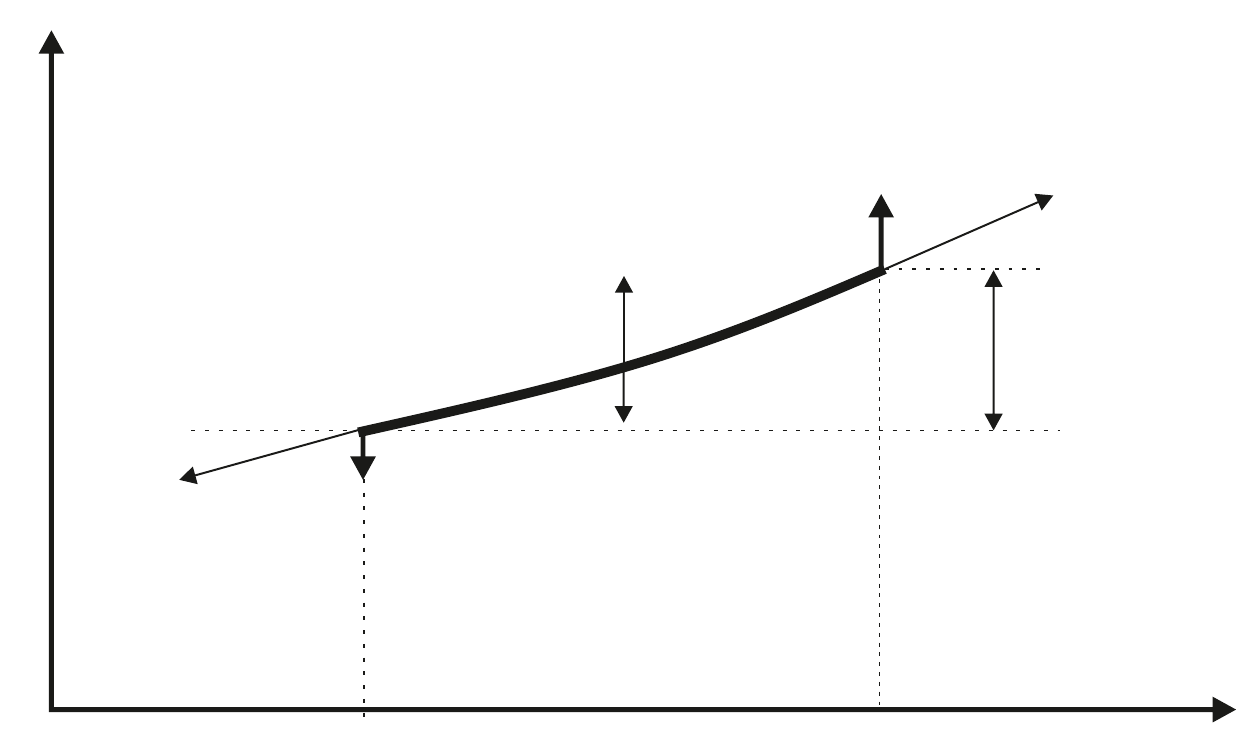}
\put(-55,15) {$ T \sin \alpha$ } 
\put(-23,37) {$ T \sin \beta$ } 
\put(-56,-2) {$z$ } 
\put(-23,-2) {$z+\ttd z$ } 
\put(-67,18) {$\alpha$}
\put(-14,23) {$\ttd u$ } 
\put(-14,31) {$\beta$} 
\put(-39,29) {$F(t)$}
\put(-39,20) {$mg$}
\put(-76,48) {$u$}
\end{picture}
\russ\\
\caption{\label{stringt}Local deformation of a taut string.} 
\end{figure}
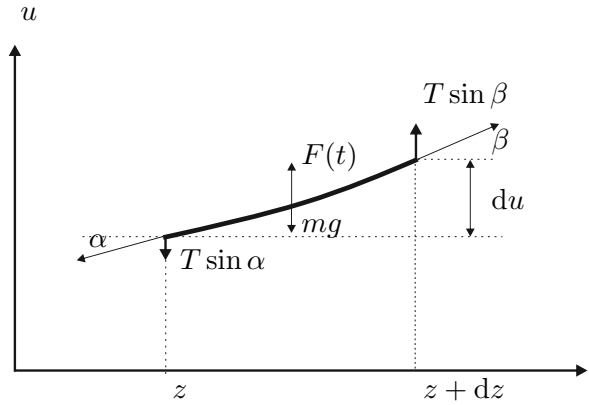

Consider a line segment $\ttd z$ in the $uz$-plane as shown in Fig.~\ref{stringt}. By the law of conservation of transverse momentum, the total force on 
the  string element must be balanced by its inertia. The net transverse force due to the difference of tension at both ends of the element is
\bq T \sin \beta - T \sin \alpha \, , \eq 
where
\bq \sin \alpha = \left. \frac{\frac{\partial u}{\partial z}}    {\sqrt{1 + \left( \frac{\partial u}{\partial z} \right)^2 }} \right|_z \, , 
\russ 
\sin \beta = \left. \frac{\frac{\partial u}{\partial z}}    {\sqrt{1 + \left( \frac{\partial u}{\partial z} \right)^2 }} \right|_{z+\ttd z}
\, . \eq
For small angles $\alpha$ and $\beta$ the horizontal forces cancel, leaving a net force in the $u$-direction. 
The tension force on the string $T$ is acting in $z$-direction and is the same at both vertices, because the segment is assumed not
to accelerate in $z$-direction.
The arc length $s(z,t)$ of the string from $0$ to $z$ is given by
\bq s(z,t) = \int_0^z  \sqrt {1 + \left( \frac{\partial u}{\partial z} \right)^2  } \ttd z = z \left( 1 + O \left( \frac{\partial u}{\partial z} \right)^2  \right) \, . \eq 
As the string length, and hence the tension, is unchanged with an error of $O \left( \frac{\partial u}{\partial z} \right)^2$, the tension is taken as a constant.
In the $u$-direction the forces on the segment can then be approximated as
\bq T \sin \beta - T \sin \alpha \approx  T \left( \left. \left( \frac{\partial u}{ \partial z}\right) \right|_{z+\ttd z} - \left. \left( \frac{\partial u }{ \partial z} \right) \right|_{z}  \right) \, . \eq
The force due to gravity is $- m g \, \vec{e_y} \cdot \vec{e_u}$ and the Lorentz force is $F(t) = I(t) B_\n{n}(z)$, where $B_\n{n}(z)$ is the~magnetic flux density normal to the~$uz$-plane (notice the typesetting of the index n).\par
The linear damping force is $- \alpha (\partial u / \partial t)$ acting in the opposite direction 
of the velocity. This is a~low-strain approximation as otherwise a term due to hysteresis damping (an energy dissipation in the~bulk of the of the string) 
would need to be considered. 
The acceleration of the string segment is $\partial^2 u / \partial t^2$. If the mass per unit length of the string is $\lambda_\n{m}$,
the inertia of the element is $\lambda_\n{m} \frac{\partial^2 u }{ \partial t^2} \ttd z$. Momentum conservation requires that
\bq \lambda_\n{m} \frac{\partial^2 u }{ \partial t^2} \ttd z = 
T \left( \left. \left( \frac{\partial u}{\partial z} \right) \right|_{z+\ttd z} \! \! \! \! \! -  \left. \left( \frac{\partial u}{\partial z} \right) \right|_{z} \right) - \lambda_\n{m} g \, \vec{e_y} \cdot \vec{e_u} \, \ttd z + 
I(t) B_\n{n}(z) \ttd z - \alpha \frac{\partial u }{ \partial t} \ttd z  \, . \label{start}  \eq
For any smooth function $f$ the Taylor series expansion yields
\bq f(z + \ttd z) - f(z) = \left( \frac{\partial f}{\partial z} \right) \ttd z + O (\ttd z)^2 \, . \eq
Disregarding the sag due to gravity\footnote{This is motivated by measuring the peak-to-peak oscillation amplitudes instead of the displacements $u(z)$.}, 
dividing by $\ttd z$ and rearranging terms in Eq.~(\ref{start}) yields the linear equation of motion 
\begin{equation}
  \lambda_\n{m} \frac{\partial^2 u }{\partial t^2}  +\alpha \frac{\partial u }{\partial t} - T \frac{\partial^2 u}{\partial z^2} = -I(t) B_\n{n}(z) , \label{dampf}
\end{equation}
where $\lambda_\n{m}$ is the mass per unit length $[\lambda_\n{m}] = 1 \, \n{kg \, m}^{-1}$,  $\alpha$ the damping coefficient $[\alpha] = 1  \, \n{kg \, m}^{-1} \, \n{s}^{-1}$. 
Note that the physical unit of the coefficient  $T/\lambda_\n{m}$ is $[T/\lambda_\n{m}] = 1 \, \n{s}^{-2}$ and 
therefore \bq c := \sqrt{\frac{T}{\lambda_\n{m}}} \eq  is the
characteristic (wave) velocity of the system. Equation (\ref{dampf}) is the dynamic force-balance equation at each point of the structure and at each 
instant of time. It has the form
\bq \mathrm{inertia + damping + spring = excitation \, force} \, . \eq 
Neglecting the damping term and the excitation function results in the scalar wave equation in one spacial dimension:
\bq \frac{\partial^2 u }{\partial t^2}  = c^2 \frac{\partial^2 u }{\partial x^2} \, . \eq 
Plane-wave eigenmodes for the free, undamped system can be found by means of the separation of the~space and time variables, 
\bq u(z,t) =  U(z) Q(t) \, . \eq 
This yields
\bq U(z)  \frac{\partial^2 Q(t) }{\partial t^2} = c^2 \frac{\partial^2 U(z) }{\partial z^2} Q(t) \, , \eq 
or
\bq \frac{ \frac{\partial^2 U(z) }{\partial z^2} }{U(z) }  =  \frac{1}{c^2}  \frac{ \frac{\partial^2 Q(t) }{\partial t^2} }{Q(t)}  \, . \eq 
The left-hand side depends only on $z$ and the right-hand side only on $t$. Both sides are equal and therefore, equal to a constant value $\mu$, which is called
an eigenvalue. Thus we obtain the two equations
\begin{align}
\frac{\partial^2 U(z) }{\partial z^2}  + \mu U(z) &= 0 \, , \label{sepp1} \\
\frac{\partial^2 Q(t) }{\partial t^2}  + c^2 \mu Q(t) &= 0  \label{sepp2} \, . 
\end{align}
In this way we have obtained the eigenvalue problems
\bq L \, U = \mu U , \russ  {\tilde L} \, Q = \mu Q \, . \eq 
The following eigensolutions fulfill Eq. (\ref{sepp1}) for $\mu < 0$:
\begin{align} 
U(z) &= {\cal A}  e^{\sqrt{-\mu} z} + {\cal B}  e^{-\sqrt{-\mu} z} \, , \\
Q(t) &= {\cal C} e^{c \sqrt{-\mu} t} + {\cal D} e^{-c \sqrt{-\mu} t} \, ,
\end{align}
for $\mu > 0$:
\begin{align} 
U(z) &= {\cal A} \sin (\sqrt{\mu} z) + {\cal B} \cos (\sqrt{\mu} z) \, , \label{eigen1}\\
Q(t) &= {\cal C} \sin (c \sqrt{\mu} t) + {\cal D} \cos (c \sqrt{\mu} t) \, ,
\end{align}
and for $\mu = 0$:
\begin{align} 
U(z) &= {\cal A} + {\cal B} z \, , \\
Q(t) &= {\cal C} + {\cal D} t  \, .
\end{align}
Only one of the three sets of solutions will be able to satisfy the boundary conditions, which separate in the same way as the displacement 
function. Substituting $u(z,t) =  U(z) Q(t)$ into the boundary conditions $u(0,t) = 0$ and $u(L,t) = 0$ yields
\bq U(0) Q(t) = U(L) Q(t) = 0 \, , \eq 
which yields $U(0) = U(L) = 0$, because otherwise $Q(t)$ would need to be zero for all times. Substituting $z=0$ into Eq. (\ref{eigen1}) 
implies ${\cal B} = 0$, substituting $z = L$ yields
\bq  {\cal A} \sin (\sqrt{\mu} z) = 0 \, , \eq 
which requires either ${\cal A} = 0$ or 
\bq \sqrt{\mu} = \frac{n \pi}{L} \, ,  \russ n=1,2,\dots . \eq 
Upon inspection, it is clear that substituting the boundary conditions into the two other sets of solutions only yield the trivial results for
${\cal A} = {\cal B} = {\cal C} = {\cal D} = 0$. 
Next we substitute the eigenvalues into Eq. (\ref{sepp2}) to find $Q(t)$. This yields
\bq 
\frac{\partial^2 Q(t) }{\partial t^2}  + c^2   \left( \frac{n \pi}{L} \right)^2  Q(t) = 0 \, ,
\eq 
with the solutions
\bq
Q_n (t) = {\cal A}_n \sin(\omega_n t) + {\cal B}_n \cos(\omega_n t) \, , 
\eq 
or 
\bq
Q_n (t) = {\cal U}_n  \sin(\omega_n t + \varphi_n) \, , 
\eq 
where 
$\omega_n$ are the normal mode frequencies
\bq  \omega_n = \frac{n \pi}{L} \sqrt {  \frac{T}{\lambda_m} } \, ,  \label{normmod} \eq
and ${\cal U}_n$ and $\varphi_n$ are still unknown coefficients to be determined from the initial conditions.  \par
Thus we obtain an infinite summation of the corresponding mode-shape functions and nodal displacement function:
\bq
  u(z,t)= \sum_n \, {\cal U}_n \sin \left(\frac{n\pi}{L}z \right) \sin(\omega_n t + \varphi_n)  \, , \label{eq:ansatz}
\eq
where the terms ${n \pi}/{L}$ are the spatial 
frequencies of the wave, which are known as {\em wave numbers}.\par
In the linearized theory, periodic solutions (normal modes) remain periodic solutions if all the~displacements are increased at the same ratio. 
Consequently, if a norm on the displacement is introduced, for example, the potential energy, then one can find a periodic solution for any prescribed value of this norm. This is  not the case for a non-linear string motion. \par
A more general form of the one-dimensional wave equation is given by 
\bq 
a^2 \frac{\partial^2 u }{\partial z^2}  + F(z,t) =  \frac{\partial^2 u}{\partial t^2}  + \alpha \frac{\partial u }{\partial t} + k u \, , 
\eq 
in the form
\bq \mathrm{spring + excitation = inertia + damping + elasticity } \, . \eq 
A well know avatar of this equation is the telegraph equation that describes the voltage $U(x,t)$ in a~segment of transmission line with resistance $R'$, inductance $L'$,
capacitance $C'$ and conductance $G'$, all defined per unit length and, in general, as a function of $x$.  $ [ R' ] = 1 \, \Omega \n{m}^{-1}$. 
\bq  
\frac{1}{L'C'} \frac{\partial^2 U }{\partial x^2}  =  \frac{\partial^2 U}{\partial t^2}  + \left( \frac{G'}{C'} + \frac{R'}{L'} \right) \frac{\partial U }{\partial t} + \frac{G'R'}{L'C'} U \, . 
\eq 
For the damped taut string ($k = 0$), separation of variables yields 
\bq \frac{ \frac{\partial^2 U(z) }{\partial z^2} }{U(z) }  =  \frac{1}{c^2}  \frac{ \frac{\partial^2 Q(t) }{\partial t^2}  +  \alpha \frac{\partial Q(t) }{\partial t}  }{Q(t)} = - \mu  \,  , \eq
or written in two equations
\begin{align}
\frac{\partial^2 U(z) }{\partial z^2}  + \mu U(z) &= 0 \, , \label{seppd1} \\
\frac{\partial^2 Q(t) }{\partial t^2}  + \alpha \frac{\partial Q(t) }{\partial t}   +   c^2 \mu Q(t) &= 0  \label{seppd2} \, . 
\end{align}
Equation (\ref{seppd2}) is a special form of a Sturm-Liouville equation with a self-adjoint differential operator. The underlying theory states that:
\begin{itemize}
\item all the eigenvalues are real
\item to each eigenvalue there is one and only one linearly dependent eigenfunction
\item the eigenfunctions are orthogonal
\item if the function is continuously differentiable on the interval, the function can be expressed as a~convergent series of these eigenfunctions.
\end{itemize}
Finding the eigenvalues and their corresponding eigenfunctions of the (spatial) boundary-value problem is a familiar task known from above. The nodal displacement function
has a different kind of solution, depending on the roots of its characteristic equation. To find the solution for $Q(t)$ we substitute $Q = e^{rt}$, which yields
\bq r^2 +  \alpha r + c^2 \mu = 0 \, . \eq 
Solving this quadratic equation for $r$ gives
\bq r = - \frac{\alpha}{2}  \pm \sqrt{  \left( \frac{\alpha}{2} \right)^2 - c^2 \mu } \, . \eq 
The solutions $Q_n (t)$, corresponding to the eigenvalues $\mu_n = \left( \frac{n \pi}{L} \right)^2$, are oscillatory if $r$ has a complex part, that is, 
if $ \left( \frac{\alpha}{2} \right)^2 - c^2 \mu_n < 0 $ for all $n$ and therefore
\bq \left( \frac{\alpha}{2} \right)^2 < \n{min} \{ c^2 \mu_n \} = \n{min} \left\{ c^2 \left( \frac{n \pi}{L} \right)^2 \right\} \, , \eq
from which follows
\bq \frac{c 2 \pi}{\alpha L} > 1 \, . \eq 
\pagebreak
Under this condition, which in practice means a "sufficiently" small damping coefficient, it holds that 
$ \sqrt{  \left( \frac{\alpha}{2} \right)^2 - c^2 \mu } = i \sqrt{  c^2 \mu - \left( \frac{\alpha}{2} \right)^2 } $ for 
$c^2 \mu - \left( \frac{\alpha}{2} \right)^2 > 0$, and $Q_n (t)$ is a linear combination of $e^{rt}$ with
\begin{align}
Q_n (t) &=   e^{- \frac{\alpha}{2} t}  \left( {\cal C}_n e^{i \sqrt{  c^2 \mu - \left( \frac{\alpha}{2} \right)^2 } t } 
                                              + {\cal D}_n e^{-i \sqrt{  c^2 \mu - \left( \frac{\alpha}{2} \right)^2 } t} \right)   \nonumber \\
&=   e^{- \frac{\alpha}{2} t}  \left( ({\cal C}_n +  {\cal D}_n ) \cos  \left( c^2 \mu - \left( \frac{\alpha}{2} \right)^2  t  \right) \right) \nonumber \\ 
& \russ \russ \left( + i   ({\cal C}_n -  {\cal D}_n )  \sin  \left( c^2 \mu - \left( \frac{\alpha}{2} \right)^2  t  \right) \right) \nonumber \\
&=   e^{- \frac{\alpha}{2} t}  \left( {\cal A}_n  \cos  \left( c^2 \mu - \left( \frac{\alpha}{2} \right)^2  t  \right) + 
                                     {\cal B}_n  \sin  \left( c^2 \mu - \left( \frac{\alpha}{2} \right)^2  t  \right) \right) 
\end{align}
or in amplitude-phase notation
\bq
Q_n (t) = {\cal U}_n  e^{- \frac{\alpha}{2} t} \sin( {\tilde \omega}_n t + \varphi_n) \, , 
\eq 
where 
\bq  {\tilde \omega}_n = \frac{n \pi}{L} \sqrt {  \frac{T}{\lambda_m} - \left( \frac{\alpha}{2} \right)^2} \, . \eq
Notice that the normal mode frequency is decreased for positive damping $\alpha > 0$ and that the normal mode decays with time, as expected.
Thus we obtain an infinite summation of the corresponding mode-shape functions and nodal displacement function:
\bq
  u(z,t)= \sum_n \, {\cal U}_n e^{- \frac{\alpha}{2} t} \sin \left(\frac{n\pi}{L}z \right) \sin( {\tilde \omega}_n t + \varphi_n)  \, . \label{eq:erg}
\eq
or
\bq
  u(z,t)= \sum_n \,  e^{- \frac{\alpha}{2} t} \sin \left(\frac{n\pi}{L}z \right) \left( {\cal A}_n \sin( {\tilde \omega}_n t)  + {\cal B}_n \cos( {\tilde \omega}_n t) \right)\, . \label{eq:erg1}
\eq
With the separation of variables method we have obtained an infinite number of solutions to the associated ordinary differential equations and their boundary conditions. 
We will now show that if we are able to represent these (known) boundary conditions in the form of a series of these solutions, the (unknown) coefficients in
these solutions can be obtained by a simple comparison of coefficients. 
We first recall that the trigonometric functions form an orthonormal basis of the Hilbert space $L^2 (\Omega), \Omega = [-\pi,\pi]$ of square integrable functions:
\begin{align}
\int_{-\pi}^{\pi} \cos(mx)\, & \cos(nx)\, dx = \frac{1}{2}\int_{-\pi}^{\pi} \cos((n-m)x)+\cos((n+m)x)\, dx = \pi \delta_{mn},   \\
 \int_{-\pi}^{\pi} \sin(mx)\, &\sin(nx)\, dx = \frac{1}{2}\int_{-\pi}^{\pi} \cos((n-m)x)-\cos((n+m)x)\, dx = \pi \delta_{mn},   \\
\int_{-\pi}^{\pi} \cos(mx)\, &\sin(nx)\, dx = \frac{1}{2}\int_{-\pi}^{\pi} \sin((n+m)x)+\sin((n-m)x)\, dx  = 0 \, . 
\end{align}
where $\delta_{mn} $ is the Kronecker delta and $m, n \ge 1$. Furthermore, the sines and cosines are orthogonal to the constant function. 
The final task is to satisfy the initial conditions of the string at rest or at a plucked position, e.g., a triangular initial profile
\bq  u(z,0) = u_0(z) \, . \eq 
Substituting the general solution (\ref{eq:erg1}) into this expressions yields 
\bq
 u_0(z) = \sum_n \, {\cal B}_n \sin \left(\frac{n\pi}{L}z \right)   \, , \label{boun1} 
 \eq 
because the exponential term is one. 
Multiplying both sides of Eq. (\ref{boun1}) 
with $\sin \left(\frac{m \pi}{L} z \right)$ and integrating over the interval $[0,L]$ gives
\bq \int_0^L \sin \left(\frac{m\pi}{L}z \right) \left(  \sum_{n} {\cal B}_n \sin \left(\frac{n\pi}{L}z \right)  \right)  \ttd z = \int_0^L \sin \left(\frac{m\pi}{L}z \right) u_0(z) \ttd z \, .\eq 
Hence
\bq 
{\cal B}_n  = \frac{2}{L} \int_0^L u_0(z) \sin \left(\frac{n\pi}{L}z \right) \ttd z = C_n \, ,
\eq 
because $\int_0^L \sin \left(\frac{m\pi}{L}z \right) \sin \left(\frac{n\pi}{L}z \right) \ttd z = L/2$ for $m=n$. The ${\cal B}_n$ are thus determined by the (known)  Fourier coefficients $C_n$ of the initial profile of the stretched wire. 
\subsection{The inhomogeneous wave equation}
The solution of the homogenous equation corresponds to the transient behavior  of the string, which depends on the initial displacement and velocity. The second 
part of the solution is the particular solution for the forcing function, which determines the steady state behavior of the string.
The gravitational force on the wire and the resulting sag give rise to a constant inhomogeneity of the equation of motion
\begin{equation}
  \lambda_\n{m} \frac{\partial^2 u }{\partial t^2}  +\alpha \frac{\partial u }{\partial t} - T \frac{\partial^2 u}{\partial z^2} = -  \lambda_\n{m} g  \, .
\end{equation}
As the right-hand side is independent of time, a constant displacement from the reference trajectory can be expected:
\begin{equation}
 T \frac{\partial^2 u}{\partial z^2} = \lambda_\n{m} g  \, .
\end{equation}
Integrating twice and considering the boundary conditions yields
\bq u(z) =   \frac{ \lambda_\n{m} g }{ 2 T } z (z - L) \, , \eq
which describes the shape of a parabola.  At  the point $z = L/2$, the function takes its minimum, which is the sag 
\bq S = \frac{\lambda_\n{m} g}{2 T} L^2 \, . \eq 
Combining this result with Eq. (\ref{normmod}) for the fundamental mode frequency yields
\bq S = \frac{g}{32}  \left(\frac{2 \pi}{\omega_1}\right)^2 \, , \eq
which allows us to determine the sag from the measurement of the fundamental frequency. 
Consider now the time dependence of the driving current expressed as $I_0 \sin(\omega t )$ so that the harmonic forcing term will be
\begin{equation}
 F(z,t) = -B_\n{n}(z) I_0 \sin(\omega t ) \, .
\end{equation}
If the forcing is sinusoidal, the response will be also sinusoidal, but it would be expected that both the~amplitude and the phase will differ from the input signal, so 
the response will be
\bq u(x,t) = \sum_n  {\cal U} \sin \left(\frac{n \pi}{L}z \right) \sin(\omega t  - \varphi_n) \, . \eq 

\begin{figure}[h]
\centering
\def\capfrac{1}
\includegraphics[clip=,width=6cm]{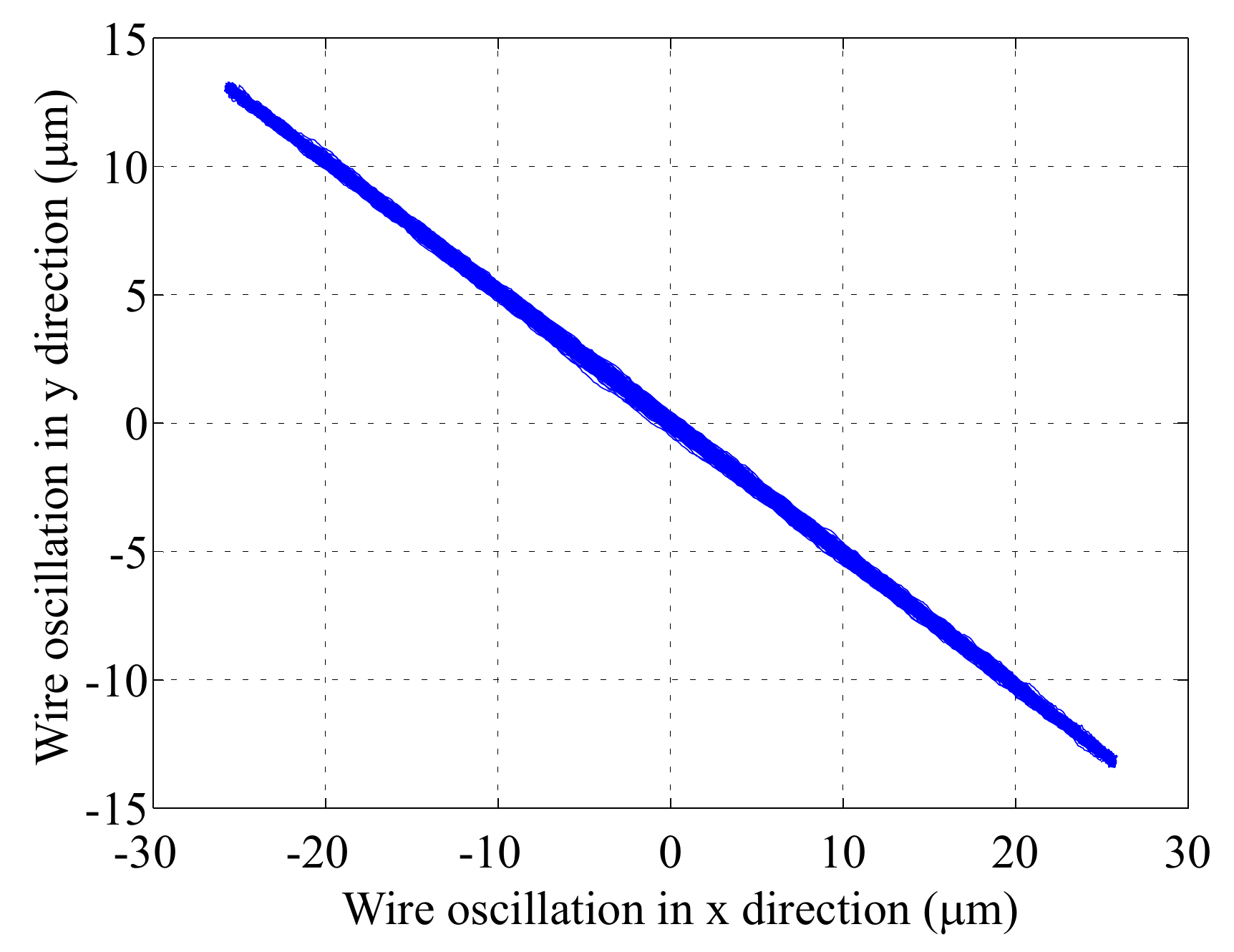}
\includegraphics[clip=,width=6cm]{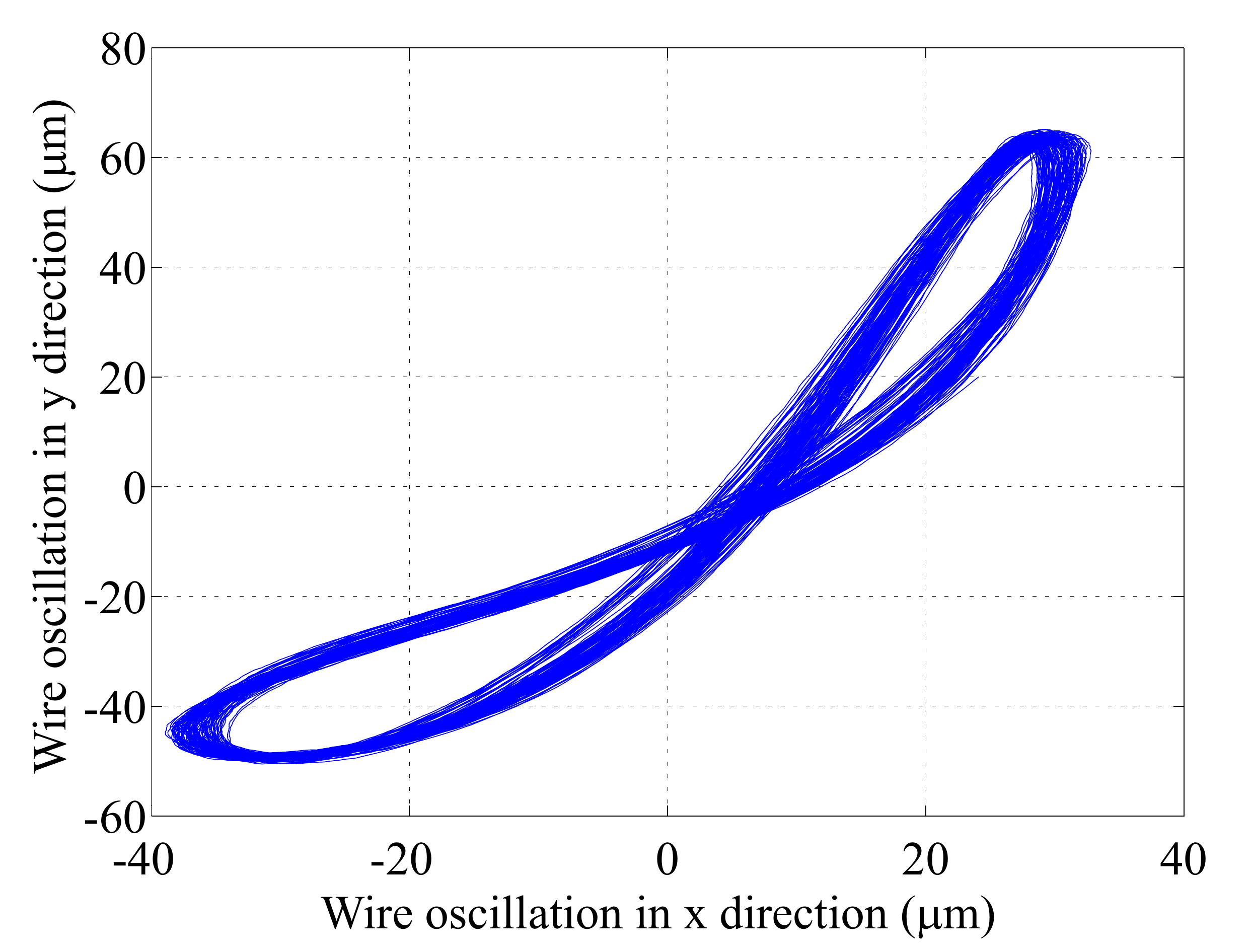}\\
\caption{\label{osci-plane} Wire oscillations in the xy-plane: Left: Measurement of a single quadrupole magnet. For the 
small oscillation amplitudes, the motion is confined to a plane. Right: Measurement of a string of magnets consisting of a quadrupole and a sextupole of about 
equal strength. The different field directions in the magnets give rise to circular polarized motion of the wire. } 
\end{figure}

\pagebreak
The method used for finding a particular solutions is an expansion of the forcing term into these spatial eigenfunctions. 
Substituting the above equation into Eq.~(\ref{dampf}) yields for $k =0$:
\begin{align}
& \sum_n {\cal U}_n  \sin \left(\frac{n \pi}{L}z \right) 
 \left( -\lambda_\n{m} \omega^2 \sin(\omega t - \varphi_n )  + \alpha \omega \cos(\omega t - \varphi_n)  + T \left( \frac{n \pi}{L} \right)^2 \sin(\omega t - \varphi_n) \right) \nonumber \\
 & \russ  = -B_\n{n}(z) I_0 \sin(\omega t ) \, . 
\end{align}
The orthogonality of the sine function can be used to eliminate the summation. Multiplying both sides 
with $\sin \left(\frac{m \pi}{L} z \right)$ and integrating over the interval $[0,L]$ yields
\begin{align}
& {\cal U}_m \frac{L}2 \left( -\lambda_\n{m} \omega^2 \sin(\omega t - \varphi_m)  +\alpha \omega \cos(\omega t - \varphi_m)  + T \left( \frac{m \pi}{L} \right)^2 \sin(\omega t - \varphi_m) \right) \nonumber \\
&  \russ \russ \label{e5} = -I_0 \sin(\omega t ) \int_0^L B_\n{n}(z)\sin \left(\frac{m \pi}{L} z \right) \mathrm{d}z \, . 
\end{align}
Equation \ref{e5} can be rearranged to 
\begin{align}
  & \left( -\lambda_\n{m} \omega^2  + T \left( \frac{m \pi}{L} \right)^2 \right)\sin(\omega t - \varphi_m) +\alpha \omega \cos(\omega t - \varphi_m) \nonumber \\
  & \russ \russ  = -\frac{2 I_0}{L {\cal U}_m} \sin(\omega t ) \int_0^L B_\n{n}(z)\sin \left(\frac{m \pi}{L} z \right) \mathrm{d}z \, .  
\end{align}
To determine the phase shift $\varphi_m$ and the coefficients ${\cal U}_m$, we use the trigonometric identity 
\bq a \sin ( \omega t - \varphi_m) + b \cos (\omega t - \varphi_m) = \sqrt{a^2 + b^2} \sin \left(\omega t - \varphi_m + \arctan \frac{b}{a} \right) \, . \eq
The $\varphi_m$ and ${\cal U}_m$ are therefore given by
\bq
\varphi_m = \arctan \left( \frac{\alpha \omega}{ -\lambda_\n{m} \omega^2  + T \left( \frac{m \pi}{L} \right)^2 } \right) \, , 
\eq
\bq
{\cal U}_m =  \frac{2 I_0}{L}  \frac{\int_0^L B_\n{n}(z)\sin \left(\frac{m \pi}{L} z \right) \mathrm{d}z}{\sqrt{\left[ T \left( \frac{m \pi}{L} \right)^2  -\lambda_\n{m} \omega^2 \right]^2 + (\alpha \omega )^2}} \, , \label{uhh}
\eq
and the particular solution for the inhomogeneous wave equation is
\begin{align}
I(t) &= I_0 \sin(\omega t)  \, , \\
u(z,t) &= \frac{2 I_0}{L} \sum_m \frac{ \int_0^L B_\n{n}(z)\sin \left(\frac{m \pi}{L} z \right) \mathrm{d}z}{\sqrt{\left[ T \left( \frac{m \pi}{L} \right)^2  -\lambda_\n{m} \omega^2 \right]^2 + (\alpha \omega )^2}} \sin \left(\frac{m\pi}{L}z \right) \sin(\omega t - \varphi_m)  \, . \label{eq:66}
\end{align}
Finally, we shall do some notational cleanup by defining the modal force as 
\bq F_m := \int_0^L I_0 B_\n{n}(z)\sin \left(\frac{m \pi}{L} z \right) \mathrm{d}z \, ,  \label{modal1} \eq 
the mode-shape function as 
\bq Y_m (z) := \sin \left(\frac{m\pi}{L}z \right) \, , \label {modal2} \eq 
the nodal displacement function as 
\bq q_m (t) := \sin(\omega t - \varphi_m) \, , \label{modal3} \eq  and
the resonance condition
\bq \omega_m = 2 \pi f_m = \frac{m \pi}{L} \sqrt { \frac{T}{\lambda_\n{m}}} =  \frac{m \pi}{L} c \label{modal4} \, . \eq
With these conventions we can write Eq.~(\ref{eq:66} ) in the compact form
\bq
u(z,t)= \frac{2}{L} \sum_k \frac{ F_m   \, Y_m (z)  \, q_m (t)  }{\sqrt{\left[ \lambda_\n{m} (\omega_m^2 - \omega^2) \right]^2 + (\alpha \omega )^2}}   \, . \label{ta68}
\eq
The phase angle is given by
\bq \varphi_m (\omega) = 
\begin{cases}
\arctan \left( \frac{\alpha \omega}{ \lambda_\n{m} ( \omega_m^2  - \omega^2 ) } \right) \, , \riss & \text{if} \quad F_m Y_m(z) > 0 \\
\arctan \left( \frac{\alpha \omega}{ \lambda_\n{m} ( \omega_m^2  - \omega^2 ) } \right) + \pi \, , & \text{if} \quad F_m Y_m(z) < 0  \, .
\end{cases}
\eq
One of the main assumptions was that the string oscillates in a plane. This is not always the case in real world situations, but it can easily be 
verified during the measurements. Two examples are shown in Fig.~\ref{osci-plane}.\par
\section{\label{fhs}Field harmonics}
In the aperture of an accelerator magnet, free of currents and
magnetized material, both magnetic scalar and vector potentials can
be employed for the formulation of a boundary value problem. For magnets that 
are long compared to their fringe-field region, a 2D field description of the integrated
field quantities is often sufficient. In 3D, the solutions are based on Fourier-Bessel series
that lead to so-called pseudo mutlipoles or generalized gradients. The treatment of 
these exceeds the aim of this introductory lecture. \par
In 2D, both formulations yield a scalar Laplace equation for
the magnetic scalar potential and for the $z$-component of the
magnetic vector potential. In what follows, we denote the problem domain
$\Omega$ and consider smooth scalar fields $\phi_\n{m}, \, A_z
\in {\cal S}(\Omega)$.

The field quality in accelerator magnets is conveniently described
by a set of Fourier coefficients, known as \emph{field harmonics}.
The method used for the calculation of field harmonics is based on
finding a general solution that satisfies the Laplace equation in a
suitable coordinate system. The integration constants in the general
solution, obtained with the separation of variables technique,
are then determined by comparison with the boundary values; in
circular coordinates these boundary values are given by the~radial
or azimuthal field components at a given
reference radius.

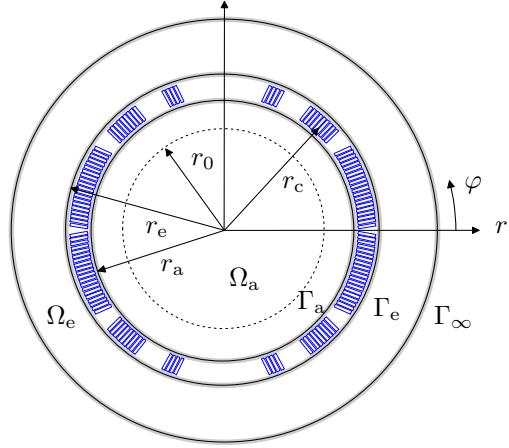
\begin{figure} [htb!]
\setlength{\unitlength}{1mm}
\begin{center}
\begin{picture}(65,60)
\includegraphics[clip=,width=6.5cm]{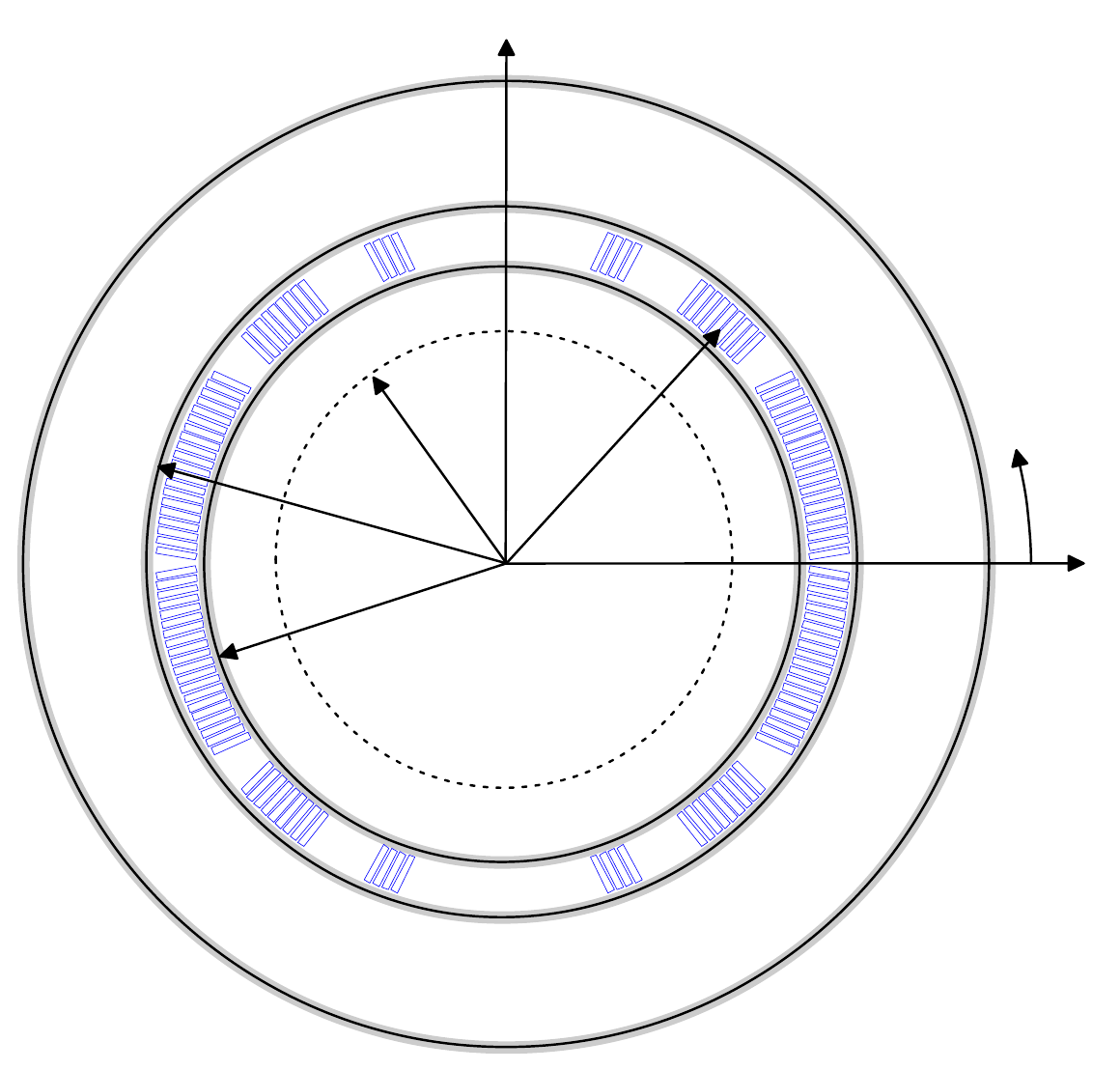}
\put(0,29){\small $r$} \put(-4,35){\small $\varphi$}
\put(-28,35){\small $r_\n{c}$} \put(-40,38){\small $r_0$}
\put(-46,29){\small $r_\n{e}$} \put(-44,24){\small $r_\n{a}$}
\put(-35,22){\small $\Omega_\n{a}$} \put(-59,17){\small
$\Omega_\n{e}$} \put(-26,19){\small $\Gamma_\n{a}$}
\put(-16,18){\small $\Gamma_\n{e}$} \put(-8,17){\small
$\Gamma_\infty$}
\end{picture}
\caption{\label{Harmdomain}Boundary value problem in 2D circular
coordinates.  The aperture domain is denoted $\Omega_\n{a}$ and
the exterior domain $\Omega_\n{e}$. The boundary $\partial \Omega_\n{a}$ 
of the aperture domain is denoted $\Gamma_\n{a}$. The boundary 
of the exterior domain consists of two parts: $\partial \Omega_\n{e}  = \Gamma_\n{e} \cup
\Gamma_\n{\infty}$.}
\end{center}
\end{figure}

In the case of accelerator magnets, the domain boundary is often chosen as a circle
with a radius of two-thirds of the aperture radius. For the time
being we shall assume that the field components are known from
measurements or numerical field calculations.
A general solution that satisfies the Laplace equation, $\nabla^2 A_z
= 0$, can be found by the separation of variables method. For
$A_z = \rho(r) \phi(\varphi)$ we obtain
\bq
\frac{\partial A_z}{\partial r } = \frac{\ttd \rho(r)}{\ttd r} \, \phi(\varphi), \russ
\frac{\partial^2 A_z}{\partial r^2} = \frac{\ttd^2 \rho(r)}{\ttd r^2} \, \phi(\varphi), \russ
\frac{\partial^2 A_z}{\partial \varphi^2} = \frac{\ttd^2
\phi(\varphi)}{\ttd \varphi^2} \, \rho(r).
\eq
Therefore, the Laplace equation in circular coordinates, 
\begin{equation}
\label{cyl222}
r^2
\frac{\partial^2 A_z}{\partial r^2} + r \frac{\partial
  A_z}{\partial r} + \frac{\partial^2 A_z}{\partial \varphi^2} = 0  \, ,  
\end{equation}
can be rewritten in the form
\begin{equation}
\label{separ} \underbrace{ \frac{1}{\rho(r)}  \left(  r^2
\frac{\ttd^2 \rho(r)}{\ttd r^2} + r \frac{\ttd \rho(r)}{\ttd r}
\right) }_{n^2} = \underbrace{ - \frac{1}{\phi(\varphi)} \,
\frac{\ttd^2 \phi(\varphi)}{\ttd \varphi^2}}_{n^2}. 
\end{equation}
Since the left-hand side of Eq.~(\ref{separ}) depends only on $r$
and the right-hand side only on $\varphi$, a separation
constant $n^2$ can be introduced, and for the case $n \neq 0$, two
ordinary differential equations are obtained:
\begin{align}
r^2 \frac{\ttd ^2 \rho(r)}{\ttd  r^2} + r \frac{\ttd  \rho(r)}{\ttd  r} - n^2 \rho(r) &= 0 , \\
\frac{\ttd ^2 \phi(\varphi)}{\ttd  \varphi^2} + n^2 \phi(\varphi) &=
0 , 
\end{align}
with the solutions,
\bq
\rho_n(r) = {\cal A}_n \, r^n + {\cal B}_n \, r^{-n} , \russ
\phi_n(\varphi) = {\cal C}_n \, \sin n\varphi + {\cal D}_n \, \cos
n\varphi . 
\eq
As the vector potential is single-valued, it must be a periodic
function in $\varphi$ with $A_z(r,0) = A_z (r,2\pi)$. The separation
constant $n$ takes integer values and therefore the general solution
of the homogeneous differential Eq. (\ref{cyl222}) is 
\begin{equation}
\label{cylsol} A_z(r,\varphi) = \sum_{n=1}^{\infty} ({\cal A}_n r^n
+ {\cal B}_n r^{-n} ) ({\cal C}_n \sin n\varphi + {\cal D}_n \cos
n\varphi ) . 
\end{equation}
We can now express the field in the aperture of an accelerator
magnet according to Eq.~(\ref{cylsol}). We will define our model problem
as shown in Fig.~\ref{Harmdomain}, consisting of an aperture
domain $\Omega_\n{a}$ and an exterior domain $\Omega_\n{e}$, with
$\partial \Omega_\n{a} = \Gamma_\n{a}$ and $\partial \Omega_\n{e} =
\Gamma_\n{e} \cup \Gamma_\n{\infty}$. 

Let us consider the magnet aperture as the problem domain. 
The condition that the flux density is finite at $r
= 0$ imposes ${\cal B}_n = 0$. In the exterior domain $\Omega_\n{e}$
the field must vanish at $\Gamma_\infty$, which results in ${\cal
A}_n = 0$. As the coefficient are not determined at this 
stage, we save on notation and rewrite 
Eq.~(\ref{cylsol}) for the vector potential in $\Omega_a$:
\begin{equation}
\label{solll} A_z(r,\varphi) = \sum_{n=1}^{\infty} r^n ({\cal
A}_n \sin n\varphi + {\cal B}_n \cos n\varphi ) . 
\end{equation}
The field components can then be expressed as
\begin{align}
\label{br} B_r(r,\varphi) &= \frac{1}{r} \frac{\partial
A_z}{\partial
  \varphi} = \sum_{n=1}^{\infty} n r^{n-1} ( {\cal A}_n \cos n\varphi
- {\cal B}_n \sin n\varphi ) , \\
\label{bf} B_{\varphi}(r,\varphi) &= - \frac{\partial A_z}{\partial
r} = - \sum_{n=1}^{\infty} n r^{n-1} ({\cal A}_n \sin n\varphi +
{\cal B}_n \cos n\varphi ), 
\end{align}
in $\Omega_\n{a}$. Each value of the integer $n$ in the solution of
the Laplace equation corresponds to a specific flux distribution
generated by ideal magnet geometries. The three lowest values, $n$ =
1,2,3 correspond to the~dipole, quadrupole, and sextupole flux
density distributions. \par
The same expressions for the field components are obtained by separating the 
Laplace equation for the magnetic scalar potential and arranging the coefficients such that
\begin{equation}
\label{sollls} \phi_\n{m} (r,\varphi) = - \sum_{n=1}^{\infty}  \frac{r^n}{\mu_0} ({\cal
A}_n \cos n\varphi - {\cal B}_n \sin n\varphi ) . 
\end{equation}
The field harmonics ${\cal
A}_n$ and ${\cal B}_n$, are undetermined at this stage. 
Assuming that the radial component of the magnetic flux density is
measured or calculated at a reference radius $r = r_0$ as a function
of the~angular position $\varphi$, we obtain the Fourier series
expansion of the radial field components:
\begin{equation}
  \label{Four1} B_r(r_0,\varphi) =
\sum_{n=1}^{\infty} ( B_n(r_0) \sin n\varphi + A_n(r_0) \cos
n\varphi
),  
\end{equation}
where
\begin{align}
A_n(r_0) &= \frac{1}{\pi} \int_{0}^{2\pi}
B_r(r_0,\varphi) \cos n\varphi \, \n{d}\varphi, \russ n=1,2,3,\ldots , \\
 B_n(r_0) &= \frac{1}{\pi} \int_{0}^{2\pi} B_r(r_0,\varphi)
\sin n\varphi \, \n{d}\varphi, \russ \, n=1,2,3,\ldots . 
\end{align}
Because the magnetic flux
density is divergence free it yields that $A_0 = 0$.
Comparing the coefficients in Eqs.~(\ref{br}) and (\ref{Four1}) we
obtain for $r_0 < r_\n{a}$:
\begin{equation}
\label{an1} {\cal A}_n = \frac{1}{n \, r_0^{n-1}} A_n (r_0) \, ,   \russ
{\cal B}_n = \frac{-1}{n \, r_0^{n-1}} B_n (r_0) \, . 
\end{equation}
The vector potential in $\Omega_\n{a}$ can now be expressed 
as a function of the multipoles 
obtained from the Fourier series expansion of the 
radial flux density, measured or calculated at $r_0$:
\begin{equation}
\label{azseries}
A_z (r,\varphi) = - 
\sum_{n=1}^{\infty}  \frac{r_0}{n}
 \left(\frac{r}{r_0}\right)^{n} (B_n (r_0) \cos n\varphi - A_n (r_0)
\sin n\varphi) .
\end{equation}
Therefore, the field components at any radius in $\Omega_\n{a}$ are given by
\begin{align}
\label{radcoil1} B_r(r,\varphi) &= \sum_{n=1}^{\infty}
\left(\frac{r}{r_0}\right)^{n-1} \! (B_n (r_0) \sin n\varphi + A_n
(r_0) \cos n\varphi)  \\  
\label{radcoil}
B_{\varphi} (r,\varphi) &= \sum_{n=1}^{\infty} \!
\left(\frac{r}{r_0}\right)^{n-1} (B_n (r_0) \cos n\varphi - A_n
(r_0) \sin n\varphi) 
\end{align}
Transformation of the components and application of De Moivre's theorem for trigonometric 
functions, yields:
\begin{align}
\label{radcoil3} B_x(r,\varphi) &= \sum_{n=1}^{\infty}
\left(\frac{r}{r_0}\right)^{n-1} \! (B_n (r_0) \sin  (n-1) \varphi + A_n
(r_0) \cos (n-1) \varphi )  \\  
\label{radcoil4}
B_y (r,\varphi) &= \sum_{n=1}^{\infty} \!
\left(\frac{r}{r_0}\right)^{n-1} (B_n (r_0) \cos (n-1) \varphi - A_n
(r_0) \sin (n-1) \varphi) 
\end{align}
The normal and skew multipole coefficients $B_n (r_0),
A_n (r_0)$ are given in units of tesla at a reference radius $r_0$.
It is common practice to normalize the multipole coefficients with respect
to the main field component\footnote{This is $B_1$ for
the dipole, $B_2$ for the quadrupole, etc.} $B_N (r_0)$, which yields in case
of the radial field component:
\begin{equation}
B_r(r,\varphi) =  B_N \sum_{n=1}^{\infty} \!
\left(\frac{r}{r_0}\right)^{n-N} (b_n (r_0) \sin n\varphi + a_n
(r_0) \cos n\varphi ) \, . 
\end{equation}
In numerical field computation, it is useful to perform a
Fourier analysis of the vector potential on the reference radius,
thus avoiding the calculation of the flux density by means of
numerical differentiation. Fourier series expansion of the magnetic
vector potential at a reference radius $r_0$ yields 
\begin{equation}
\label{azentwick}
A_z (r_0,
\varphi) = \frac{C_0}{2} + \sum_{n=1}^{\infty}  (C_n (r_0) \sin n\varphi + D_n (r_0) \cos n\varphi)  \, . 
\end{equation}
Substituting (\ref{an1}) into Eq.~(\ref{solll}), setting $C_0 = 0$ and comparing the coefficients, we obtain
\begin{equation}
\label{mpoletran}
B_n (r_0) = \frac{-n \, D_n (r_0)}{r_0}
\, , \russ A_n (r_0) = \frac{n \, C_n (r_0)}{r_0} \, . 
\end{equation}
For the magnetic scalar potential we obtain
\begin{equation}
\phi_\n{m} (r,\varphi) = - \sum_{n=1}^{\infty} \frac{r_0}{n \mu_0} \left(\frac{r}{r_0}\right)^{n} (A_n (r_0) \cos n\varphi + B_n (r_0) \sin n\varphi ) . 
\end{equation}
The curl operator acting on vector fields has thus been replaced by algebraic operations
in the corresponding sequence space $l^2$ of the Fourier series. A similar procedure applies for the
series expansion of the magnetic scalar potential. \par
Hence we can express the components of the magnetic flux density as a function of the 
multipoles obtained from the Fourier series expansion of the vector potential at $r_0$:
\begin{align}
B_r(r,\varphi) &= \sum_{n=1}^{\infty} \frac{n}{r_0} \, 
\left(\frac{r}{r_0}\right)^{n-1} \! (- D_n (r_0) \sin n\varphi + C_n
(r_0) \cos n\varphi )  , \\ 
B_{\varphi} (r,\varphi) &= \sum_{n=1}^{\infty} \frac{-n}{r_0} \, 
\left(\frac{r}{r_0}\right)^{n-1} (D_n (r_0) \cos n\varphi + C_n
(r_0) \sin n\varphi). 
\end{align}
Instead of the radial component of the magnetic flux density we may expand the azimuthal and the~Cartesian components along the closed
circle inside the aperture of the magnet, that is, the $2\pi$ periodic signals $S = B_r, B_\varphi, B_x, B_y, A_z, \phi_\n{m}$ are transformed 
using the discrete Fourier transform and expressed as 
\bq S = \sum_{n=1}^{\infty}  (C_n (r_0) \sin n\varphi + D_n (r_0) \cos n\varphi) \,. \eq
Comparison of coefficients yields the results shown in Table \ref{develop}. \par
\begin{table}[t]
\begin{center}
 \caption{\label{develop}Relations between the multipole coefficients and the Fourier coefficients of the expansion of $B_r, B_\varphi, B_x, B_y$, and $A_z$, $\phi_\n{m}$.}
\renewcommand{\arraystretch}{1.3}
\setlength{\tabcolsep}{15pt}
 \begin{tabular}{lcccccc}\hline \hline
                   &   \pmb{$B_r $}       &    \pmb{$B_\varphi$}    &   \pmb{$B_x$}          &    \pmb{$B_y$}    & \pmb{$A_z$}  & \pmb{$\phi_\n{m}$} \\  \hline
$B_n$ =        &   $C_n$ &   $D_n$  &  $C_{n-1}$ & $D_{n-1}$  &  $\frac{-n D_n}{r_0}$ & $\frac{- n \mu_0 C_n}{r_0 }$  \\
$A_n$ =        &   $D_n$ &   $-C_n$ &  $D_{n-1}$ & $-C_{n-1}$ &  $\frac{n C_n}{r_0}$  &  $\frac{- n \mu_0 D_n}{r_0 }$ \\
\hline
\hline
\end{tabular}
\end{center}
\label{dB}
\end{table}
Using this technique we have: 
\begin{itemize}
\item found a convenient way to describe the field imperfections in accelerator magnets
\item have been able to reconstitute the vector field $\vec{B}$ in the entire aperture of the magnet from measurements of only one component on its circular boundary, and
\item use the systems variables of numerical field computation (magnetic vector potential and magnetic scalar potential) without the need for local, numerical differentiation.
\end{itemize}
\section{\label{vib}The oscillating-wire method}
\begin{figure}[h]
\centering
\includegraphics[clip=,width=9cm]{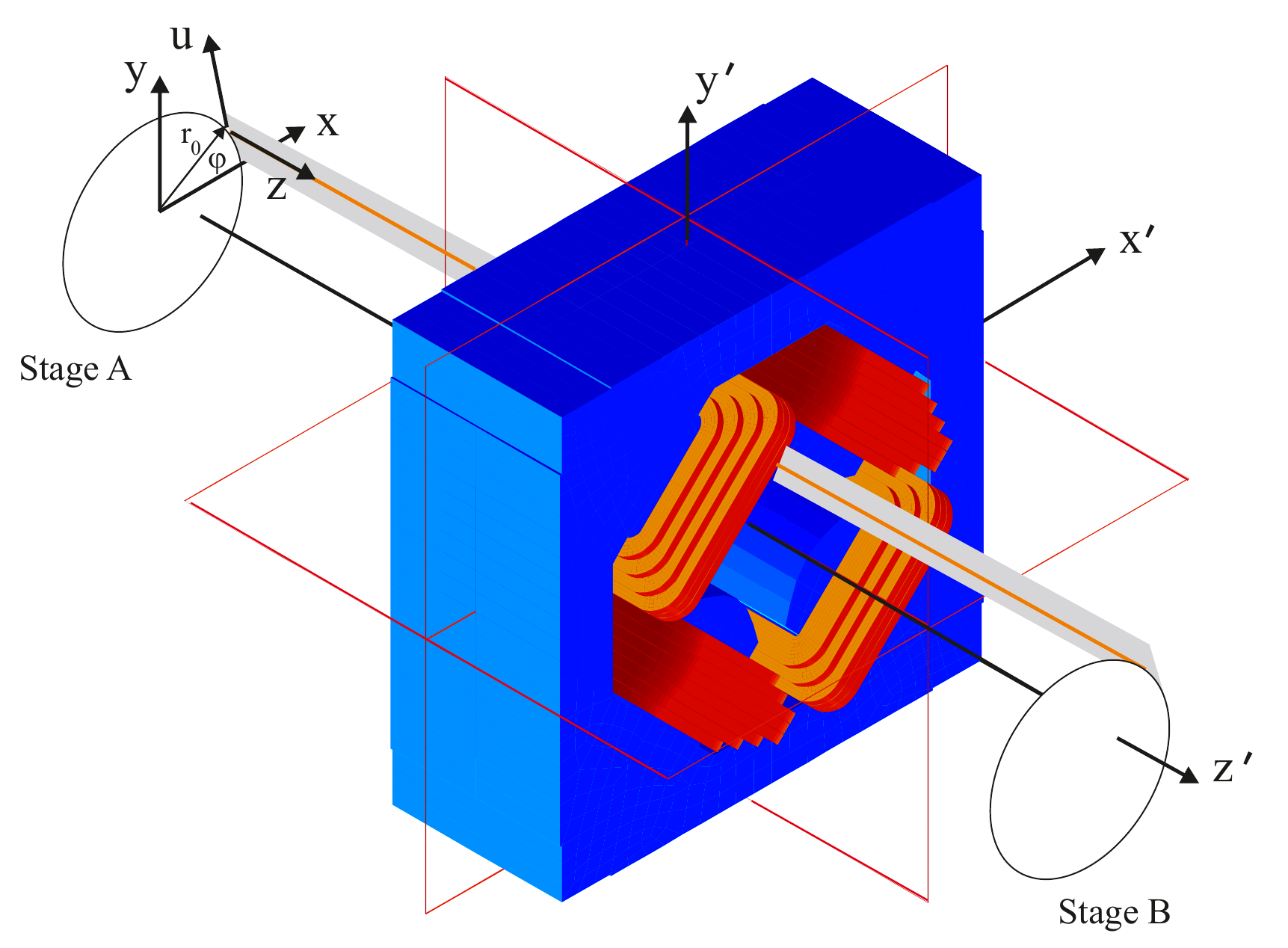}
\caption{\label{wireframe} Reference frame for magnet axis and plane of wire oscillation} 
\end{figure}

To measure the transversal field harmonics, the wire is positioned step-by-step on the generators of a~cylindrical domain inside the magnet aperture, that is, its end-points 
at the stages are moved on a~circular trajectory $(r_0, \varphi_k), k = 1, \cdots , K$ in the transverse plane. This is shown schematically in Fig.~\ref{wireframe}.  The~longitudinal coordinate
is denoted $z$, as usual. The radius $r_0$ of the circular trajectory is chosen 
as large as possible in order to increase the sensitivity of the multipole measurement. \par
The frequency of the driving current in the wire is chosen such that 
the resonance frequencies are avoided. In this way, the oscillation can be treated as a quasi-static problem 
when the settling time of the~system is respected.\par
The method is based on the Lorentz force displacement 
of the wire, which is fed by an alternating current. The wire thus vibrates in the direction orthogonal to the magnetic field. 
If the amplitude of the~wire motion is small, it can be assumed (and experimentally verified) that the wire moves in a plane with
coordinates $u$ and $z$. \par

At each position $(r_0, \varphi_k)$, the wire displacements are measured by means of phototransistors arranged in orthogonal directions at a longitudinal position $z_0$ close to Stage A. In this way, the wire displacements $d_x (r_0, \varphi_k)$ and $d_y (r_0, \varphi_k )$  are obtained. The displacements in $x-$ and $y$-directions are proportional to the Lorentz force, which is in turn proportional to the integral field components:
\bq  d_y^k (r_0) = \lambda_y  \int_0^L B_x (r_0, \varphi_k ) \ttd z \, , \riss  d_x^k (r_0) = \lambda_x  \int_0^L B_y (r_0,  \varphi_k ) \ttd z \, . \eq
Experiments have shown that the effect of gravity,  causing wire sag in $y$-direction, can be minimized by  measuring the peak-to-peak oscillation amplitudes rather than the absolute wire displacements. The~displacements $d_x^k (r_0)$ and $d_y^k (r_0)$ are then determined from the wire amplitudes $\delta_x^k (r_0) $ and $\delta_y^k (r_0)$ by correlating the voltage signal from the phototransistors with the phase of the driving current measured over a shunt resistor.\par
\bq \label{signum} d_x^k (r_0) = \delta_x^k (r_0)  \, \n{sign}  \left(\sigma_{x,k} -\frac{\pi}{2} \right) \, , \riss  d_y^k (r_0) = \delta_y^k (r_0)  \, \n{sign} \left(\sigma_{y,k}-\frac{\pi}{2} \right) \, , \eq
where the $\sigma$ are the phase angles between the voltage signals from the phototransistors 
and the excitation current.  Measuring the wire amplitudes yields a 
rectified signal for the $x$ and $y$-channels. To recconstruct $2\pi$ periodic functions in $\varphi$, the polarity must be introduced by the 
signum function in Eq.~(\ref{signum}). \par
We will now {\em assume} that the proportionality between the integrated field components and the wire displacement is independent of the angular position $\varphi_k$; a statement that we must challenge later.  Under this assumption, the Fourier series expansion can be performed directly on the wire displacements, that is,
\bq {\tilde A}_n (r_0) = \frac{2}{K} \sum_{k=0}^{K-1} d_y^k (r_0) \cos \, n \varphi_k \, , \riss  {\tilde B}_n (r_0) = \frac{2}{K} \sum_{k=0}^{K-1} d_y^k (r_0) \sin \, n \varphi_k \, . \eq
The tilde is used to distinguish these from the coefficients obtained from the series expansion of the~radial field component. However, by normalizing the Fourier coefficients 
${\tilde A}_n (r_0)$ and ${\tilde B}_n (r_0)$  to the main component, and again under the assumption that the $\lambda_x$  and $\lambda_y$ do not vary for the different angular positions, the relative multipoles can be obtained from:
\bq \label{osrel}  a_{n+1} (r_0) = \frac{ {\tilde A}_n (r_0) }{ {\tilde B}_N (r_0) } \, , \riss b_{n+1} (r_0) = \frac{ {\tilde B}_n (r_0) }{ {\tilde B}_N (r_0) } \, .\eq
The index $n+1$ stems from the fact that the wire displacement is proportional to the integrated $B_x$ component and not the $B_r$ component; note the relations
shown in Table~\ref{develop}. \par
Equation (\ref{osrel}) thus relate in a straightforward manner the relative field multipoles to the coefficients of the Fourier series of the wire oscillation 
amplitudes (in the $x-$ and $y$-directions), measured on the circular trajectory.  Owing to the holomorphic properties of the integrated field, the field harmonics calculated separately 
on the set of displacements $d_x (r_0, \varphi_k)$ and $d_y (r_0, \varphi_k)$, must yield the same results. Both $x$ and $y$ oscillation amplitudes are nevertheless acquired for 
redundancy and checking the effect of gravity and wire suspension at the stages.\par
Let now the maximum wire displacements at a given
longitudinal position $z_0$ be denoted 
\bq d_x^k := \n{max}_t \{u_x (r_0, \varphi_k, z_0,t) \}, \riss  
d_y^k := \n{max}_t \{u_y (r_0, \varphi_k, z_0,t) \} \, . \eq
If these displacements are directly proportional to the integrated 
$B_y$ and $B_x$ components, the discrete Fourier transforms of $d_x^k$ or  $d_y^k$ yield the relative transverse field harmonics 
as long as the proportionality factor $\lambda_k$
does not vary as a function of the wire's angular position.  \par
To derive the theory of the taut string moving in the $uz$-plane we consider, more generally, 
the~amplitude $u(z,t) := u (r_0, \varphi_k, z,t)$ and the static field component normal to this plane of oscillation 
$B_\n{n} (z) := B_\n{n} (r_0, \varphi_k, z)$ and omit, in what follows, the notation of the position $r_0$ and $\varphi_k$.  \par
As shown above, the steady state solution of the corresponding wave equation is given by 
\bq
u(z,t)= \frac{2 I_0}{L} \sum_m \frac{ \int_0^L B_\n{n}(z)\sin \left(\frac{m \pi}{L} z \right) \mathrm{d}z}{\sqrt{\left[ T \left( \frac{m \pi}{L} \right)^2  -\lambda_\n{m} \omega^2 \right]^2 + (\alpha \omega )^2}} \sin \left(\frac{m\pi}{L}z \right) \sin(\omega t - \varphi_m)  \, , \label{taut67}
\eq
when the wire is excited with the current $I(t)= I_0 \sin(\omega t)$.  Using again the notation for the modal force, Eq.~(\ref{modal1}),  
the mode-shape function (\ref{modal2}), the nodal displacement function (\ref{modal3}), and the resonance 
condition (\ref{modal4}) we can write
\bq
u(z,t)= \frac{2}{L} \sum_m \frac{ F_m   \, Y_m (z)  \, q_m (t)  }{\sqrt{\left[ \lambda_\n{m} (\omega_m^2 - \omega^2) \right]^2 + (\alpha \omega )^2}}  \, . \label{taut68}
\eq
Considering the wire properties, the phase angles 
for all oscillation modes are very small ($<$ 0.1 deg), that is, the wire follows the field instantaneously.  
It can be seen from Eq.~(\ref{taut68}) that the shape of the wire motion is given by a linear combination of eigenmodes that depend on the modal force, the wire length, 
the tension and the viscous-damping coefficient. The quotient of integrated field component $ \int_0^L B_\n{n}(z) \mathrm{d}z$ and oscillation amplitude at $z_0$ 
may therefore depend on the angular position $\varphi_k$ of the wire. To better understand the shape of the oscillating wire described by Eq.~(\ref{taut68}), we will first study some special cases. \newline \par
\subsection{Fourier transform of the field distribution}
The term $\int_0^L B_\n{n}(z)\sin \left(\frac{m \pi}{L} z \right) \mathrm{d}z$ can be reduced if we express the longitudinal variation of the 
magnetic flux density by a Fourier series.
\newline
\newline
\bq B_\n{n}(z)= \sum_n {\cal C}_n \sin \left(\frac{n \pi}{L} z \right) \,,  \label{longfh} \eq
where
\bq 
 {\cal C}_n= \frac{2}{L} \int_0^L B_\n{n}(z)\sin \left(\frac{n \pi}{L} z \right) \mathrm{d}z . \eq
\newline 
 We thus obtain:
 \newline
\begin{eqnarray}
  u(z) \sim \sum_m \frac{ {\cal C}_m }{\sqrt{\left[ \lambda_\n{m} (\omega_m^2 - \omega^2)  \right]^2 + (\alpha \omega )^2}} \sin \left(\frac{m\pi}{L}z \right) \, .  \label{longhh}
\end{eqnarray}
\newline
If the shape of the vibrating wire is known, the left-hand-side of this equation can be expanded into Fourier series and the (so far unknown) ${\cal C}_m$ can be obtained by a simple comparison of 
coefficients.  \par
\pagebreak
\subsection{Vanishing tension; the slackline}
We can further neglect constant factors and the time dependency.
If the wire is not tensioned ($T=0$), the denominator in the sum becomes independent of $k$ and can therefore be written in front of the sum:
\newline
\newline
\begin{equation}
 u(z) \sim  \frac{1}{\sqrt{\left( \lambda_\n{m} \omega^2 \right)^2 + (\alpha \omega )^2}} \sum_m {\cal C}_m \sin \left(\frac{m\pi}{L}z \right) = 
  \frac{1}{\sqrt{\left( \lambda_\n{m} \omega^2 \right)^2 + (\alpha \omega )^2}} B_\n{n}(z) \, . 
\end{equation}
\newline
\newline
Thus, if the wire swings loosely in the magnetic field, it will take exactly the form of the magnetic field. But since we measure the displacement only at one point $z_0$ 
outside the magnet's field region, the~measured displacement will be zero.\newline \par
\subsection{The hard-edge magnet model}
Consider a magnet with a field that can well be characterized by a hard-edge model, that is, the transversal field drops instantaneously to zero, or in other words, there is no fringe field. Magnets built with permanent magnet excitation come relatively close to this assumption.  The longitudinal distribution of the magnetic field simplifies a lot in this case. The amplitude can therefore said to be $B_0$ and the (active) length of the magnet is $2 \varepsilon$ around the center at ${L}/{2}$. Thus, the integral results in:
\newline

\begin{align}
&B_0 \int_{\frac{L}{2}-\varepsilon}^{\frac{L}{2}+\varepsilon} \sin \left(\frac{m \pi}{L} z \right) \mathrm{d}z  \nonumber =
B_0 \left[ -\frac{L}{m\pi} \cos \left(\frac{m\pi}{L}z \right) \right]_{\frac{L}{2}-\varepsilon}^{\frac{L}{2}+\varepsilon}  \nonumber \\
& \riss =  - \frac{B_0 L}{m\pi} \left[ \cos \left(\frac{m\pi}{L} \left(\frac{L}{2}+\varepsilon \right) \right) - \cos \left(\frac{m\pi}{L} \left(\frac{L}{2}-\varepsilon \right)\right) \right]  \nonumber \\
& \riss= - \frac{B_0 L}{m\pi} \left[ -2 \sin \left(\frac{m \pi}{2} \right) \sin({\varepsilon}) \right]  \, , 
\end{align}
\newline
\newline
where we have used the trigonometric identity $\cos z-\cos u=2\sin \frac{u+z}{2}\sin \frac{u-z}{2}$. This results in:

\bq
  u(z,t)= \frac{4 I_0 B_0}{\pi} \sin(\varepsilon) \! \! \! \sum_{m=1,3,5,7,...} \! \! \!  \! \! \!  \frac{\sin \left(\frac{m \pi}{2} \right) }{m}  \left( \left[ \lambda_\n{m} (\omega_m^2 - \omega^2)  \right]^2 \! + (\alpha \omega )^2 \right)^{- \frac{1}{2}}  Y_m (z) \, q_m (t)  .  \nonumber
\eq

The terms under the summation are all independent of the angular position $\varphi_k$ of the wire within the magnet bore. Therefore, the amplitude of the oscillation is strictly proportional to the transversal field component $B_0$, as required. Note, however, that the transversal field component is indeed a function of $\varphi_k$. 
\pagebreak
\subsection{Intrinsic measurement error of the oscillating-wire method}
\begin{figure}[h]
\centering
\includegraphics[clip=,width=7.5cm]{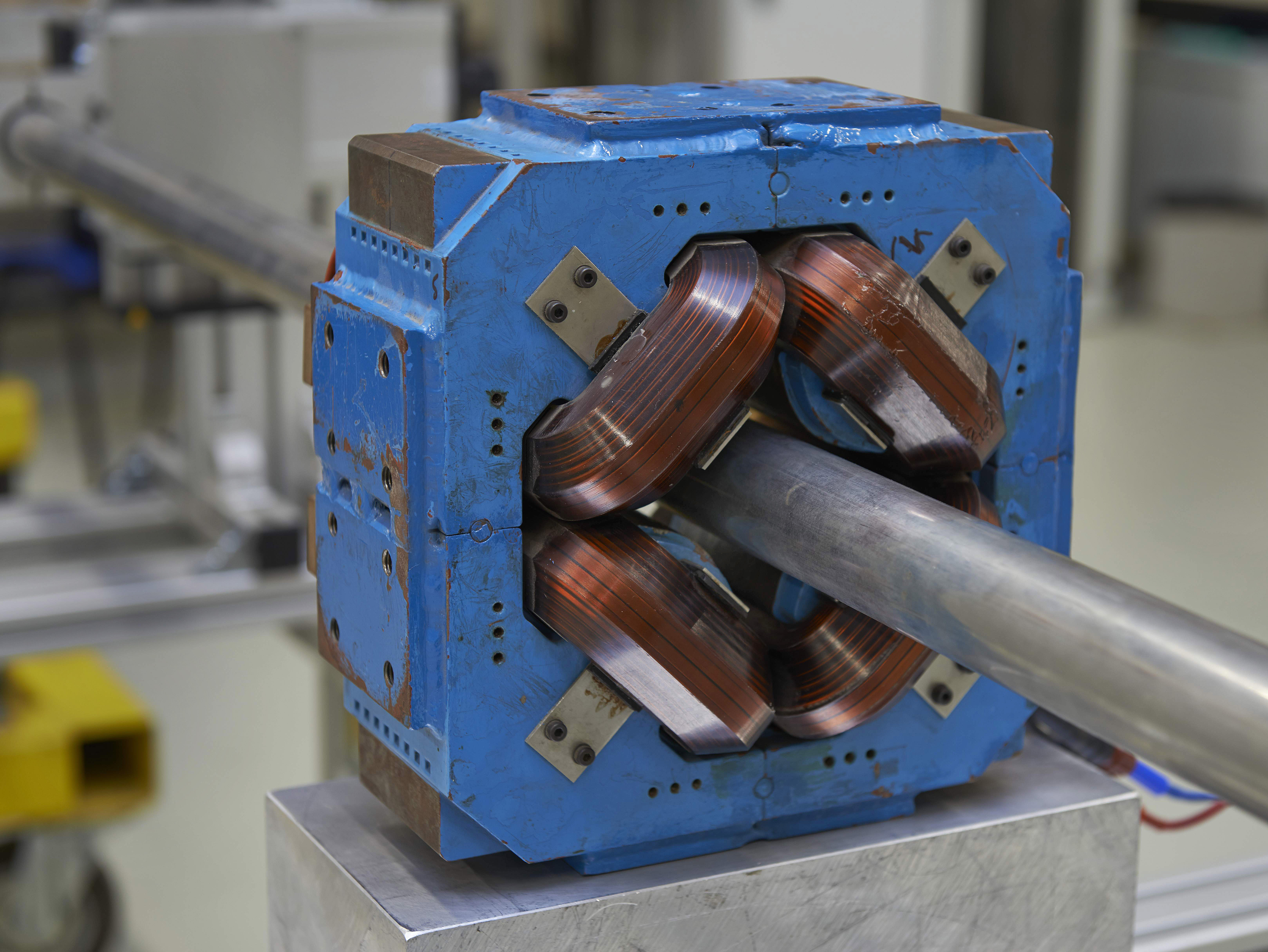}
\caption{\label{testgeom}Photograph of a reference dipole magnet with a tube mounted to accommodate a rotating-coil magnetometer. This transducer was 
used as a reference for the validation of the stretched-wire technique. The numerical model of the magnet is shown in Fig. \ref{wireframe}.}
\end{figure}

Simulations have shown that the intrinsic error made by only measuring the oscillation amplitude at a~single point, is fortunately relatively small. This error can be estimated 
with the simulated 3D field distribution of the magnet.  The numerical field computation ROXIE was used to calculate the field for the~magnet that has been measured with the oscillating-wire method. 
Simulation routines in Matlab\textsuperscript{\textregistered} were developed to study the influence of the different wire parameters on the multipoles measurements. 
The magnet is an iron-dominated quadrupole designed for the LEP project at CERN. According to the~CERN naming convention, this magnet
is subsequently referred to as the LEP-IL-QS magnet.  A spare unit  is used as a reference magnet in CERN's magnetic 
measurement laboratory.  Figure~\ref{testgeom} shows a photograph of this magnet with a tube mounted to accommodate a rotating-coil magnetometer. This transducer was 
used as a reference for the validation of the stretched-wire technique.\newline \par
\begin{figure}[h]
\centering
\def\capfrac{1}
\includegraphics[clip=,width=9.5cm]{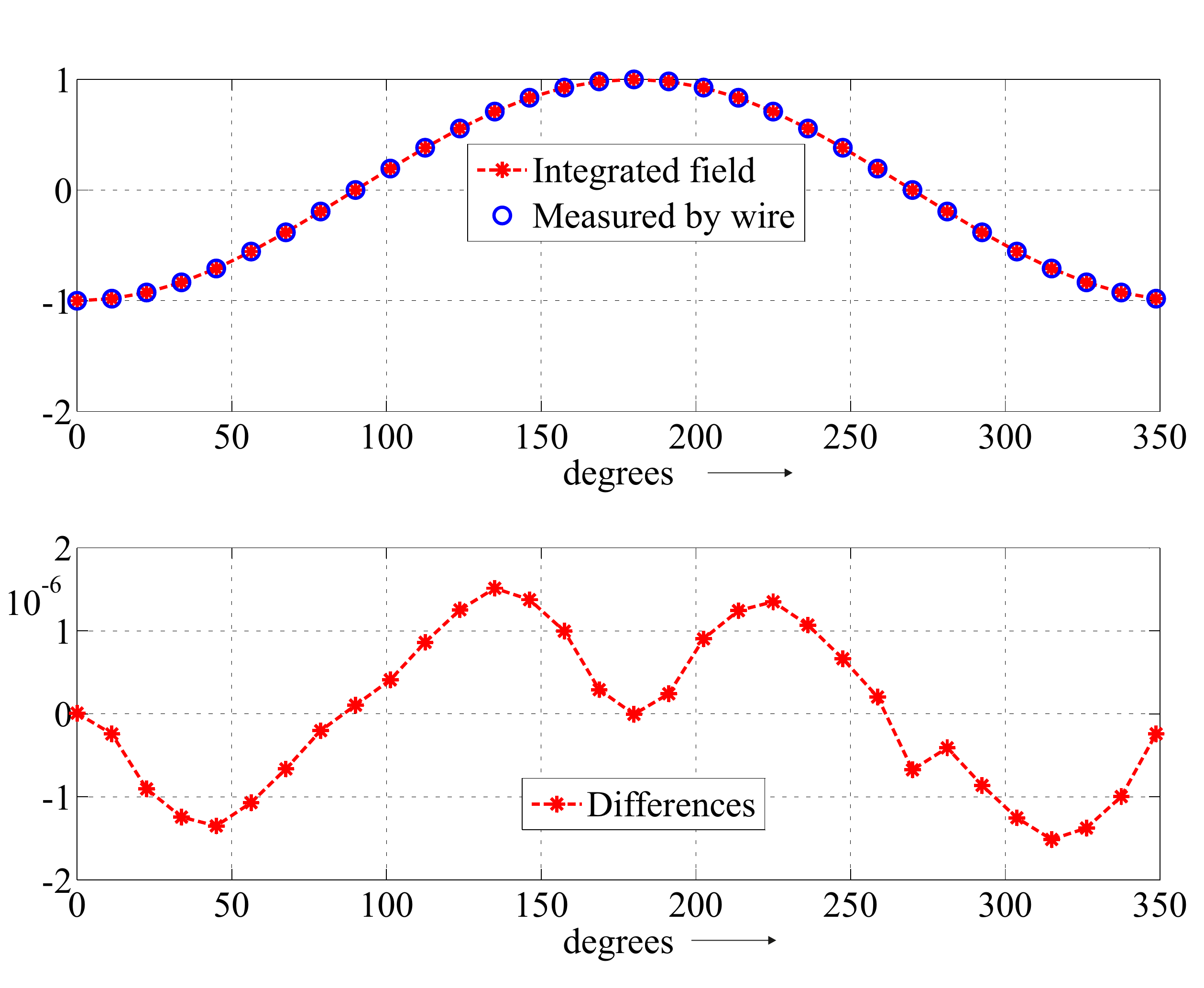}
\includegraphics[clip=,width=9.5cm]{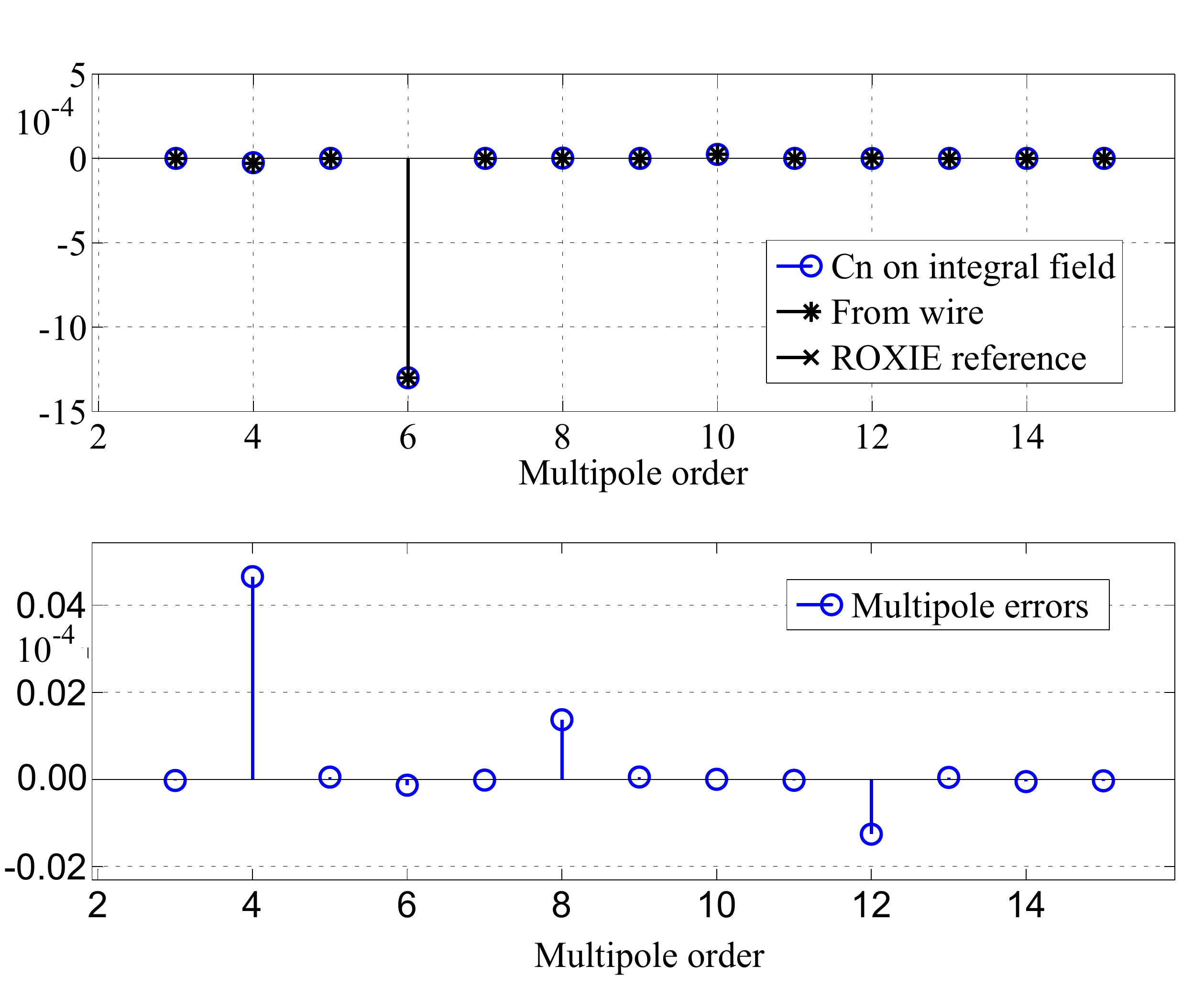}
\caption{\label{test6} Oscillation amplitudes and integrated flux densities for 32 angular positions $\varphi_k$. Top: Normalized oscillation amplitude at $z = 0.1 $m compared to the normalized quantity of the integrated magnetic flux density. Second from top: Differences of the normalized oscillation amplitudes and integrated flux densities. Second from bottom: Transversal field harmonics reconstructed from the wire oscillation amplitudes and reference values from field computation.  Bottom: Differences between reconstructed and
reference values. }
\end{figure}

For the positioning angles between zero and $2 \pi$, Fig \ref{test6} (top) shows the normalized oscillation amplitude at $z = 0.1$ m compared to the normalized quantity of the integrated magnetic flux density $\int_0^L B_{x} \ttd z$. The differences of the normalized oscillation amplitudes and integrated flux densities are given in the second image from the top. This error is on the order of a few units in $10^{-6}$.   \par

Figure~\ref{test6} also shows the field harmonics reconstructed from the wire oscillation amplitudes and reference values from field computation and rotating-coil measurements. 
Notice that for this magnet the~only relevant field error is the integrated sextupole component of about -12 units in $10^4$. The intrinsic errors (due to an assumption of constant proportionality and in-plane motion) have propagated into the~transversal $b_4$ field harmonics with about  0.04 units in $10^{4}$.  This error is small compared to the systematic measurement errors 
(from stage misalignment and systematic variations in the wire tension) and the  uncertainty errors from due to environmental effects, non-linearity in the sensors, 
and non-infinitely rigid suspension points.
\pagebreak
\newline
\pagebreak

\end{document}